\documentclass[12pt, a4paper]{article}
\usepackage{amsfonts,amsmath,amssymb,epsfig}
\usepackage{subfigure}

\newcommand{\be}{\begin{equation}}
\newcommand{\en}{\end{equation}}

\newcommand{\II}{\textrm{I}}
\newcommand{\JJ}{\textrm{J}}
\newcommand{\YY}{\textrm{Y}}
\newcommand{\KK}{\textrm{K}}
\newcommand{\DD}{\textrm{D}}
\newcommand{\dd}{\textrm{d}}

\newcommand{\filler}{\hspace*{\fill}}

\begin{document}

\title{A wave near the edge of a circular disk}

\author{
  \filler M.~Destrade, Y.B. Fu  }
\date{2008}
\maketitle

%%%%%%%%%%%%%%%%%%%%%%%%
\begin{abstract}

It is shown that in
the Love-Kirchhoff plate theory, an edge wave can travel in a 
circular thin disk made of an isotropic elastic material. 
This disk edge wave turns out to be faster than the classic flexural 
acoustic wave in a straight-edged, semi-infinite, 
thin plate, a wave which it mimics when the curvature radius 
becomes very large compared to the wavelength.

\end{abstract}
%%%%%%%%%%%%%%%%%%%%%%%%

%%%%%%%%%%%%%%%%%%%%%%%%
\section{INTRODUCTION}
%%%%%%%%%%%%%%%%%%%%%%%%

When an elastic half-space is subjected to a point excitation, it 
responds by generating waves which become surface (Rayleigh) waves in 
the far-field. 
When a cymbal is hit on its edge, it produces a long-lasting sound: 
why not conjecture that this phenomenon is due to the propagation of 
an acoustic wave localized near the edge of the cymbal?

The study of \textit{vibrations} of thin circular plates with free 
edges dates back to at least 1910 with a paper by Airey \cite{Aire10}, 
and is still a very active area of research (see for instance 
Touz\'e, Thomas, and Chaigne \cite{ToTC02}). 
In contrast there seems to be almost no contribution on the 
subject of \textit{waves} in these plates, 
whereas there exist a wealth of 
papers on waves over the surface of cylinders with infinite extent. 
That scarcity might be explained as follows. 
In the latter case (cylinder), 
the asymptotic behavior of a wave when the radius 
of the cylinder becomes very large (in comparison to the wavelength)
is well-known: it is that of the celebrated Rayleigh wave in a 
half-space \cite{Rayl85}.
In the former case (disk), 
the asymptotic behavior is not as well-known: 
it is that of a wave propagating along the straight, free edge of 
a thin Kirchhoff plate \cite{Kone60}. 
Hence \v{C}erv \cite{Cerv88} studied waves traveling 
on the surface of a thin disk, 
but used plane-stress equations, where the thickness of the plate does 
not appear: as a result the asymptotic behavior of his waves is that 
of Rayleigh waves. 

This paper shows that  
the appropriate resolution of the classical small-deflection 
equations of motion in a circular disk causes no particular problem. 
It is conducted with a view to interpret some patterns of motion 
observed near the edge of a stricken cymbal \cite{ScGR98} 
(other experimental results are found in  \cite{BeCh84}).

\begin{figure}[!t]
\centering
\epsfig{figure=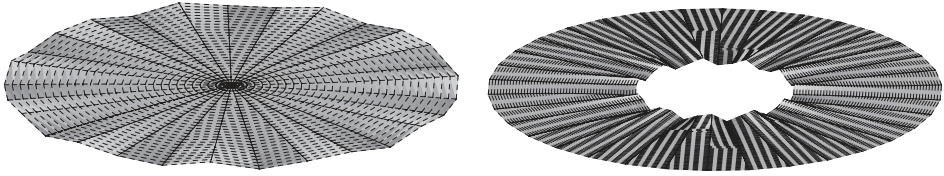,width=.9\textwidth}
 \renewcommand{\figurename}{Fig.}
 \caption{Wave localized near the edge of a disk or a hole.}
\label{fig1}
\end{figure}

Other related problems are left open but should pose no particular 
difficulty to solve; they include the following. 
(i) The complimentary problem of acoustic waves propagating along the 
edge of a circular hole in a thin plate, 
with obvious possible applications in non-destructive evaluation,
see Figure~\ref{fig1}.
(ii) A circular disk made of one material in rigid or slippery contact 
with a holed plate made of another material; 
there the asymptotic behavior is that of contact flexural waves in 
thin plates \cite{ZiSu83} and the applications are also in the realm 
of non-destructive evaluation (of the strength of rivets for example).
(iii) Waves traveling in a thin flat annulus; 
there the asymptotic behavior is that of flexural waves on a narrow 
plate \cite{Norr03} and an exotic application would be the 
transportation of planar heavy objects by acoustic 
levitation \cite{HaKU98} on circular transporting systems.
(iv) Another possibility resides in linking the present results to the 
experimental work of Shaw \cite{Shaw56} on Barium Titanate 
disks.

%%%%%%%%%%%%%%%%%%%%%%%%%
\section{BASIC EQUATIONS}
%%%%%%%%%%%%%%%%%%%%%%%%%

The equations of motion for a Love-Kirchhoff plate are \cite{Mans64}
\be
(V_s h)^2 \; \nabla^4 w + 6(1-\nu) \; \ddot{w} = 0,
\en
where $w$ is the deflection, $h$ is the thickness of the plate, 
$\nu$ its Poisson ratio, and $V_s$ its shear wave speed.
Now consider an acoustic flexural wave traveling along the 
circumference of a circle (axis normal to the plate, radius $R$) 
in the form
\be \label{wave}
w = W(r)\cos (k R \theta - \omega t),
\en
where $r$, $\theta$ are the in-plane cylindrical 
coordinates, $k$ is the wavenumber, and $\omega$ the frequency. 
Introducing the dimensionless ``speed'' $\alpha$, defined 
by \cite{ThMc74}
\be
\alpha^2 = 6(1-\nu) \left(\dfrac{V}{V_s}\right)^4
                     \left(\dfrac{V_s}{\omega h}\right)^2, 
\quad \text{where} \quad 
V:= \dfrac{\omega}{k},
\en
the equations of motion can be put in the form
\be
\DD_+ \DD_- W = \DD_- \DD_+ W = 0,
\en
where $\DD_\pm$ are the following differential operators:
\be 
\DD_\pm := \dd^2 / \dd r^2 + (1/r)  \dd / \dd r - k^2 R^2 / r^2 
              \pm \alpha k^2.
\en

Clearly, the general solution to this differential equation is a 
linear combination of Bessel functions; specifically,
\be \label{generalSolution}
W(r) = \gamma_1 \JJ_p(\sqrt{\alpha}kr)
        + \gamma_2 \II_p(\sqrt{\alpha}kr)
         + \gamma_3 \YY_p(\sqrt{\alpha}kr)
          + \gamma_4 \KK_p(\sqrt{\alpha}kr), 
\quad p:=kR.
\en
Here, the $\gamma_i$ are constants and the labeling of Bessel 
functions and modified Bessel functions follows the conventions of 
Abramowitz and Stegun \cite{AbSt65}.
In particular,
\be \label{prop}
\DD_+  \JJ_p(\sqrt{\alpha}kr) = \DD_+  \YY_p(\sqrt{\alpha}kr) = 
\DD_-  \II_p(\sqrt{\alpha}kr) = \DD_-  \KK_p(\sqrt{\alpha}kr) = 0.
\en

Then for the edge of the circle to be free, two boundary 
conditions must be satisfied \cite{Mans64}. 
First, the edge shear must vanish, so that 
\be \label{BC1}
\dfrac{\partial^2 w}{\partial r^2} + 
  \nu \left(\dfrac{1}{r} \dfrac{\partial w}{\partial r}
            + \dfrac{1}{r^2} \dfrac{\partial^2 w}{\partial \theta^2}
             \right) =0
\quad \text{at } r = R.
\en 
Second, the rate of change of the twisting moment must also vanish, 
 so that 
\be \label{BC2}
\dfrac{\partial}{\partial r} (\nabla^2 w) + 
  \dfrac{(1-\nu)}{r} 
    \dfrac{\partial}{\partial \theta}
      \left(
       \dfrac{1}{r} \dfrac{\partial^2 w}{\partial r \partial \theta}
         - \dfrac{1}{r^2} \dfrac{\partial w}{\partial \theta}
             \right) = 0
\quad \text{at } r = R.
\en 

Finally, the displacement at $\theta$ is the displacement at 
$\theta + 2 \pi$, $\theta + 4 \pi$, etc., and so by \eqref{wave}, 
$p=kR$ must be an integer. 
In other words, the wave can only travel under the condition that 
\textit{its 
wavelength is an integer fraction of the circle perimeter $2\pi R$}.

%%%%%%%%%%%%%%%%%%%%%%%%%%%%%%
\section{CIRCULAR DISK}
%%%%%%%%%%%%%%%%%%%%%%%%%%%%%%

For a wave on the edge of a circular disk, take 
$\gamma_3 = \gamma_4 =0$ in \eqref{generalSolution} to avoid 
singularities at the center of the disk. 
Then the boundary conditions \eqref{BC1} and \eqref{BC2} reduce,
after much manipulation and use of \eqref{prop}, to 
\begin{multline}
 \JJ_p(\sqrt{\alpha} p) 
 \left[(1-\nu) j_p(\alpha)/p - (1 - \nu - \alpha) \right] \gamma_1
\\
 + \II_p(\sqrt{\alpha} p) 
  \left[(1 - \nu) i_p(\alpha)/p + (1 - \nu + \alpha) \right]  \gamma_2
 = 0,
\end{multline}
and 
\begin{multline}
 \JJ_p(\sqrt{\alpha} p) 
 \left[(1 - \nu + \alpha) j_p(\alpha) - (1 - \nu)/p \right] \gamma_1
 \\  
+ \II_p(\sqrt{\alpha} p) 
 \left[ (1 - \nu - \alpha) i_p(\alpha) - (1 - \nu)/p \right] \gamma_2
= 0,
\end{multline}
respectively. 
Here the functions $j_p(\alpha)$ and $i_p(\alpha)$ are defined by 
\begin{align}
j_p(\alpha) :=
  \sqrt{\alpha} \;
    \dfrac{\JJ'_p(\sqrt{\alpha} p)}{\JJ_p(\sqrt{\alpha} p)}
\sim \sqrt{1 - \alpha} \quad \text{as } p \rightarrow \infty, 
\notag \\
i_p(\alpha) :=
  \sqrt{\alpha} \;
    \dfrac{\II'_p(\sqrt{\alpha} p)}{\II_p(\sqrt{\alpha} p)}
\sim \sqrt{1 + \alpha} \quad \text{as } p \rightarrow \infty, 
\end{align}
and their asymptotic behavior as $p$ becomes large is found from 
Debye series expansions \cite{AbSt65}.
For non-trivial solutions to exist, the following 
\textit{dispersion equation} must be satisfied:
\begin{multline} \label{disp}
 (1 - \nu - \alpha)^2 i_p(\alpha) - (1 - \nu + \alpha)^2 j_p(\alpha)
\\
 + 2 \alpha [1 + i_p(\alpha) j_p(\alpha)] (1 - \nu) / p    
  -[i_p(\alpha) - j_p(\alpha)]  (1 - \nu)^2 /p^2 = 0.
\end{multline}
Not surprisingly, the dispersion equation for disk edge waves is the 
same as the frequency equation \cite{Aire10, Leis93} derived for the 
\textit{vibrations} of a circular disk with free edges, simply  
because the governing equations and the boundary conditions happen to 
coincide in the present context for waves such as \eqref{wave} and 
for vibrations in the form 
$w = W(r)\cos (k R \theta) \cos(\omega t)$, say.
The interpretation is however different. 
In particular, when the radius of the disk becomes large with respect 
to the wavelength, $p \rightarrow \infty$ and this equation becomes 
that of the edge wave on a semi-infinite plate
discovered by Konenkov \cite{Kone60} and hereafter called 
the ``straight-edge wave'',
\be
 (1 - \nu - \alpha)^2 \sqrt{1 + \alpha} = 
     (1 - \nu + \alpha)^2 \sqrt{1 - \alpha}.
\en

Note that in general, the dispersion equation \eqref{disp} gives 
several roots, numbered by an integer parameter $s$, say. 
The smallest root ($s=0$) corresponds to a solution which is never 
zero over the surface of the disk; the next root ($s=1$) corresponds 
to a solution which is zero once over the surface of the disk, and 
so on. 
Hence $s$ denotes the number of nodal circles \cite{Leis93}.

Table 1 displays the values of $\alpha$ for various Poisson ratios 
and integer values of $p$ at $s=0$; clearly, $\alpha$ tends to the 
straight-edge wave speed as $p \rightarrow \infty$ 
(last column, see also Thurston and McKenna \cite{ThMc74});
also, the disk edge wave is always faster than the straight-edge 
wave (note that similarly, 
the wave propagating on the free surface of a 
\textit{cylinder} is faster than the Rayleigh wave propagating over 
the free surface of an elastic half-space \cite{Vikt67}.) 
For the one-mode ($p=1$) wave, $\alpha$ is nearly twice its value for
the straight-edge wave ($p=\infty$) and then it decreases rapidly as 
$p$ grows. 
Multiplying the values displayed by $p^2$ gives the quantity computed 
and collected by Leissa \cite{Leis93};
the reader is also referred to that book for values when $s \ge 1$. 

\begin{center}
Table 1.
\textit{Values of $\alpha$ for various $\nu$ and $p$ ($s=0$)}
{\small
\rule[-3mm]{0mm}{8mm}
\begin{tabular}{l|c c c c c c}
 $_\nu$ \verb!\! $^p$ & 5 & 10 & 20 & 50 & 100 & $\infty$
 \\\hline
0.1  & 1.43125 & 1.30545 & 1.20294 & 1.11354
     & 1.07218 & 0.999985
 \\
0.17 & 1.40355 & 1.28864 & 1.19284 & 1.10814 
     & 1.06862 & 0.999852
 \\
0.2  & 1.39035 & 1.28028 & 1.18757 & 1.10510
     & 1.06656 & 0.999697
 \\
0.25 & 1.36639 & 1.26460 & 1.17736 & 1.09907
     & 1.06227 & 0.999177
 \\  
0.3  & 1.33980 & 1.24654 & 1.16513 & 1.09144
     & 1.05668 & 0.998102
 \\  
1/3  & 1.32048 & 1.23302 & 1.15570 & 1.08533
     & 1.05209 & 0.996899
 \\
0.4  & 1.27763 & 1.20198 & 1.13326 & 1.07013
     & 1.04031 & 0.992639 
 \\\hline
\end{tabular}
}
\end{center}

\bigskip

For the purpose of comparison, recall that 
Viktorov \cite{Vikt67} studied waves over the convex surface of a full 
cylinder, plotted their variation with depth for a cylinder made of 
brass, at $p = 5, 41, \infty$, and found that they `decay somewhat 
more rapidly with distance from the surface than for a Rayleigh wave, 
the rate of decay being faster for smaller $p=kR$'. 
Figure 2  shows that the depth profiles (normalized with respect to 
the wavelength) at $p = 5, 41, 200, \infty$ 
and $s=0$ for a brass disk follow a similar trend.
Note however that the edge waves on a disk decay \textit{much more} 
rapidly than the straight-edge wave, which is known to be 
\textit{weakly} inhomogeneous \cite{Norr94}.
\begin{figure}[!t]
\centering
\epsfig{figure=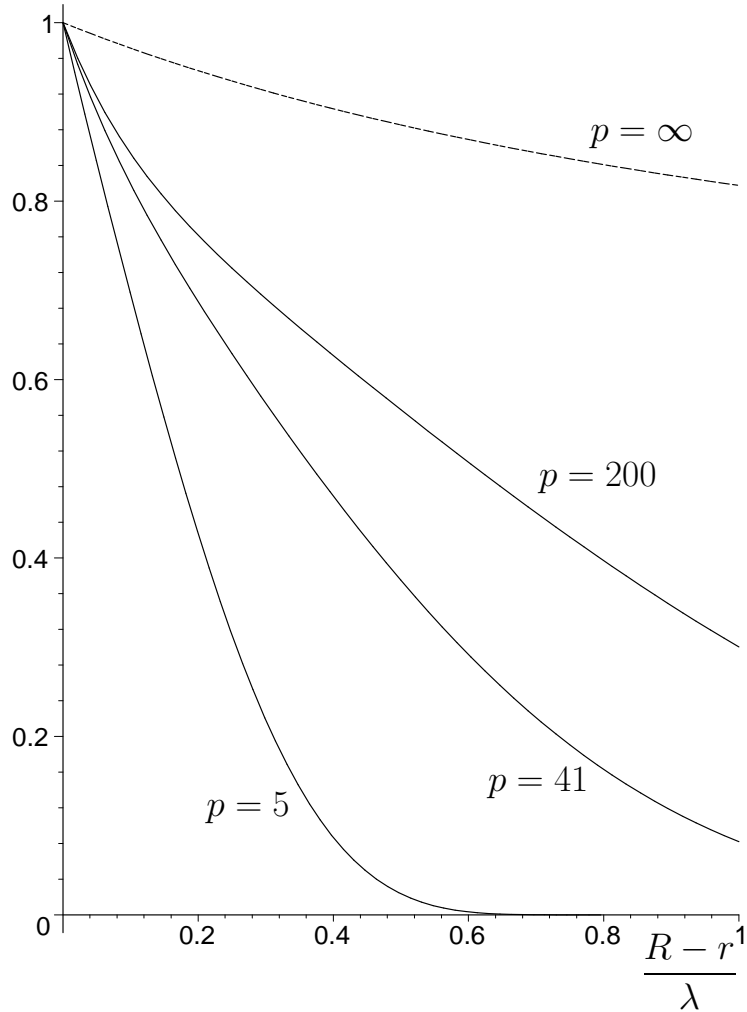,width=.7\textwidth}
 \renewcommand{\figurename}{Fig.}
 \caption{Depth profiles for a thin disk made of brass ($\nu = 0.34$).}
\end{figure}

Schedin, Gren, and Rossing \cite{ScGR98} used double-pulsed television
holography  
to record the response of a cymbal made of brass ($\nu = 0.34$) when 
it is struck near its edge. 
They noted that `high-frequency waves are apparent along the edge' 
and these can be clearly seen on their pictures. 
A rough estimate gives a number of about 12 wavelengths in a $\pi R/8$ 
sector, indicating that $p$ is about 200. 
Figure 3 displays the dispersion curves for brass for 
$s=0, 1, 2, 3,$ up to  $p=200$. 
Here, the discrete nature of the plots has been over-ridden for ease 
of reading.
\begin{figure}[!t]
\centering
\epsfig{figure=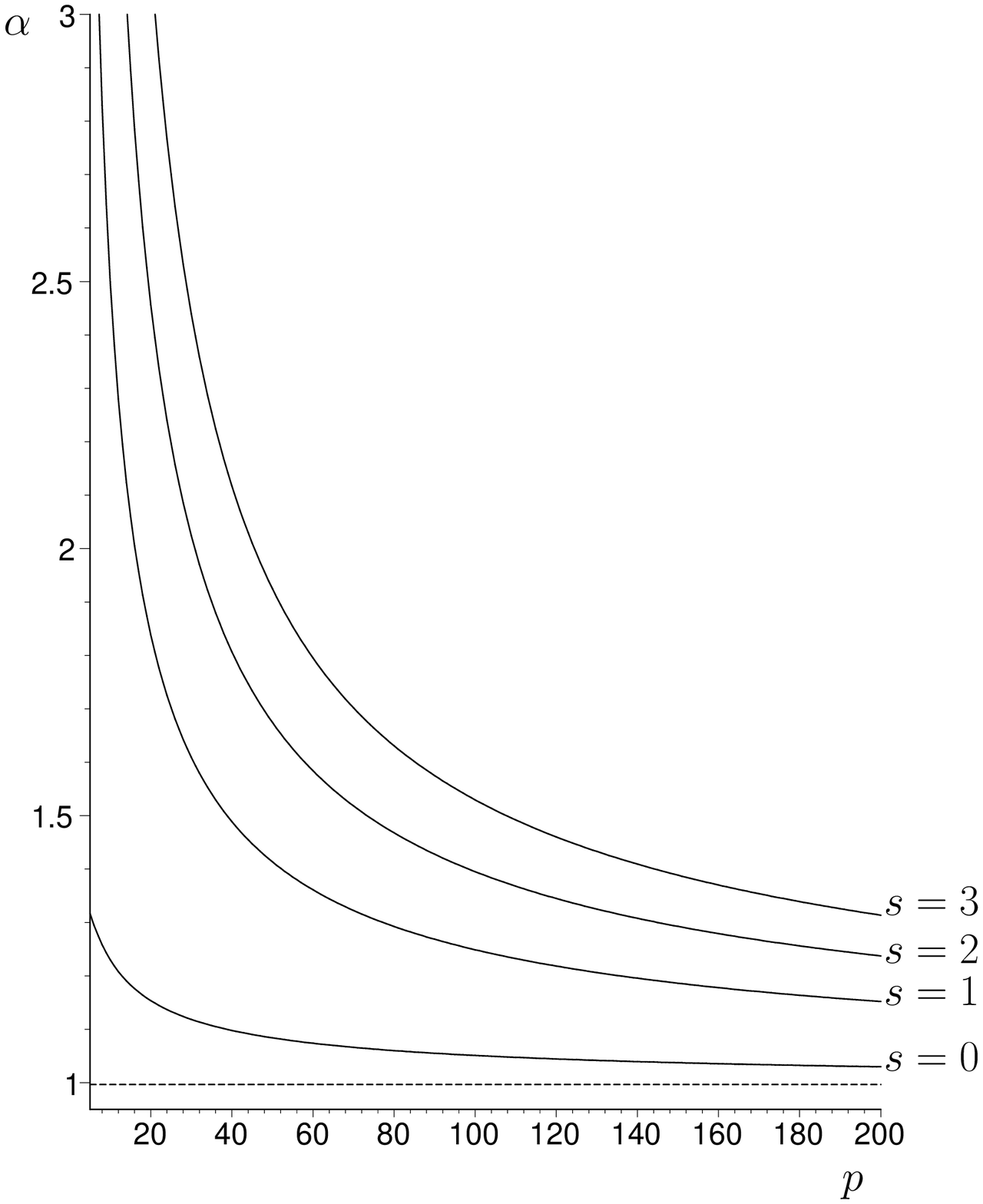,width=.7\textwidth}
 \renewcommand{\figurename}{Fig.}
 \caption{Dispersion curves for a disk made of brass ($\nu = 0.34$).
The dashed line represents the speed of the straight-edge wave.}
\end{figure}
Finally, the profiles of the disk are shown on Figure 4 at $p=200$ for 
$s=0, 1, 2$; there the distance was normalized with respect to the 
radius of the disk and the elevation at the edge of the disk was taken 
as half the radius (exaggerated scale).
For $s=0$ (no nodal circle), the amplitude decays rapidly away 
from the edge, and is almost zero at $0.9R$. 
For $s=1,2$, the amplitude of the wave reaches a maximum which is 
greater than its value at the edge (amplification); it is also 
less localized than for $s=0$, being almost zero at about $0.8R$.
Seen from above, the case $s=2$ (two nodal circles) will display 
one shaded region and one bright region close to the edge, 
a pattern which is somewhat similar to that recorded by 
Schedin, Gren, and Rossing \cite{ScGR98}.
\begin{figure}[!t]
\centering
\epsfig{figure=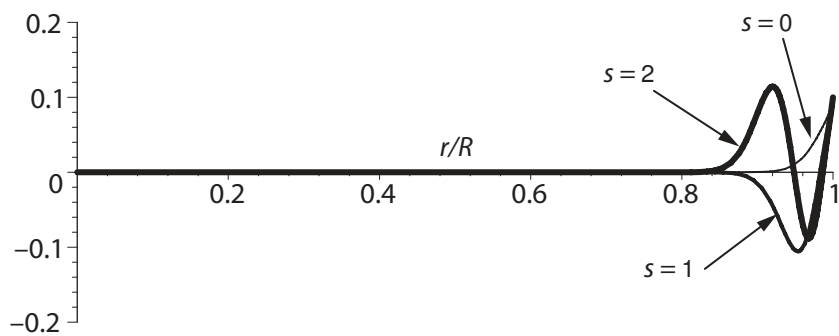,width=.8\textwidth}
 \renewcommand{\figurename}{Fig.}
 \caption{Profile of a disk made of brass ($\nu = 0.34$) when 
$p=200$, $s=0,1,2$.}
\end{figure}

%%%%%%%%%%%%%%%%%%%%%%%%
%++++++++++++++++++++++++++++++++++++++++++++++++++++++
% bibliography
%++++++++++++++++++++++++++++++++++++++++++++++++++++++

%%%%%%%%%%%%%%

\end{document}